\documentclass[10pt]{article}
\usepackage{fullpage}
\usepackage{amsmath}
\usepackage{amssymb}
\usepackage{color}
\usepackage[dvips]{epsfig}
\usepackage{cite}

\advance \textwidth by -2in
\advance \textheight by -3in
\def\red{\textcolor{red}}

\def\##1{{\bf #1}}
\def\=#1{\underline{\underline #1}}
\def\~#1{{\tilde{\bf #1}}}

\def\eps{\epsilon}
\def\epso{\epsilon_0}
\def\muo{\mu_0}
\def\ko{k_0}
\def\co{c_0}

\def\etao{\eta_0}
\def\.{\mbox{ \tiny{$^\bullet$} }}

\def\epsr{\eps_{r}}
\def\mur{\mu_{r}}

\def\curl{\nabla\times}
\def\curlp{\nabla^\prime\times}

\def\divp{\nabla^\prime\.}

\def\ro{(\#r)}
\def\rpo{(\#r^\prime)}
\def\rto{(\~r)}

\def\le{\left(}
\def\ri{\right)}
\def\les{\left[}
\def\ris{\right]}
\def\lec{\left\{}
\def\ric{\right\}}

\def\c#1{\cite{#1}}
\def\l#1{\label{#1}}
\def\r#1{(\ref{#1})}

\def\ux{\hat{\#x}}
\def\uy{\hat{\#y}}
\def\uz{\hat{\#z}}

\def\ur{\hat{\#r}}
\def\uphi{\hat{\mbox{\boldmath$\phi$}}}
\def\utheta{\hat{\mbox{\boldmath$\theta$}}}

\def\newM{\mbox{\boldmath${\cal M}$}}
\def\newN{\mbox{\boldmath${\cal N}$}}

\def\calJ{{\cal J}}
\def\calK{{\cal K}}
\def\calP{{\cal P}}
\def\calQ{{\cal Q}}
\def\calR{{\cal R}}
\def\calU{{\cal U}}
\def\calV{{\cal V}}

\def\tcalJ{\tilde{\cal J}}
\def\tcalK{\tilde{\cal K}}
\def\tcalP{\tilde{\cal P}}
\def\tcalQ{\tilde{\cal Q}}
\def\tcalR{\tilde{\cal R}}

\def\st{\sin\theta}
\def\ct{\cos\theta}
\def\sp{\sin\phi}
\def\cp{\cos\phi}

\def\sfS{{\sf S}}
\def\sfV{{\sf V}}

\def\sfSl{{\sf S_\ell}}
\def\sfVl{{\sf V_\ell}}
\def\Ul{\=U_\ell}

\def\urho{\hat{\mbox{\boldmath$\rho$}}}

\def\smn{_{smn}}

\def\adj{\mbox{adj}}

\begin{document}

\Large
\begin{center}
{\bf {Vector spherical wavefunctions for orthorhombic
dielectric-magnetic material with gyrotropic-like magnetoelectric
properties}}

 \vspace{5mm}

\normalsize

{\bf Akhlesh  Lakhtakia}$^1$\footnote{Corresponding author. E--mail: akhlesh@psu.edu}  and
{\bf Tom G. Mackay}$^{1,2}$

\vspace{2mm}

 $^1$\emph{NanoMM---Nanoengineered Metamaterials Group, Department of Engineering Science and
Mechanics, Pennsylvania State University, University Park, PA
16802--6812, USA}

 \vspace{2mm}

$^2$\emph{School of Mathematics  and
   Maxwell Institute for Mathematical Sciences, \\University of Edinburgh, Edinburgh
EH9 3JZ, United Kingdom}

\vspace{15mm}

{\bf Abstract}

\vspace{2mm}

\end{center}

\normalsize

\noindent \emph{Vector spherical wavefunctions were derived in
closed-form to represent time-harmonic electromagnetic fields
in an orthorhombic dielectric-magnetic material with
gyrotropic-like magnetoelectric properties. These wavefunctions were
used to formulate the T matrix for scattering by a three-dimensional object
composed of the chosen material.
Furthermore, a closed-form, coordinate-free expression of the dyadic Green function
for the chosen material was derived. Expressions ascertained for the singularity 
behavior will be useful for formulating volume integral equations for scattering
inside the chosen material. A bilinear expansion of the dyadic Green function was obtained
in terms of the derived vector spherical wavefunctions.
}

\vspace{2mm}


\section{Introduction}
Closed-form solutions of the vector differential   equation
\begin{equation}
\label{Eq1}
\nabla\times\left[\nabla\times{\#F}(\#r)\right]-\kappa^2{\#F}(\#r)=\#0
\end{equation}
in a spherical coordinate system to represent  the field $\#F$ and the position vector $\#r$
can be traced back to as early as the 1860s, when Clebsch
considered the scattering of an elastodynamic wave by a solid sphere, as recounted by  Logan in 1965
\cite{Logan1965}. In frequency-domain electromagnetics for homogeneous, isotropic, dielectric materials,
this equation  simplifies to the vector Helmholtz equation
\begin{equation}
\label{Eq2}
\nabla^2{\#F}(\#r)+\kappa^2{\#F}(\#r)=\#0
\end{equation}
because of  the constraint $\nabla\.{\#F}(\#r)=0$. Credit for this reduction and application to the optical scattering
response of a homogeneous, isotropic, dielectric sphere
evidently goes to a series of papers by Ludvig Lorenz from the second half of the
19th century \cite{Logan1965,Prinkey1994}. However, Lorenz's 1880 paper \cite{Lorenz1890,Lorenz1898,Kragh1991} has
largely been superseded in popular memory by the 1908 paper of Gustav Mie \cite{Mie1908}, most probably because Lorenz
used a pre-Maxwellian theory of light whereas Mie used the frequency-domain Maxwell equations and also
applied his solution to the very practical problem of  scattering of light by  colloidal spheres.

Solutions of Eqs. \r{Eq1} and \r{Eq2} in  spherical coordinates are called
vector spherical wavefunctions.
For homogeneous, isotropic, dielectric
materials, these wavefunctions involve $\kappa = \ko\sqrt{\epsr}$, where $\ko$
is the free-space wavenumber and $\epsr$ is the (frequency-dependent) relative permittivity scalar. Extension
to homogeneous, isotropic, dielectric-magnetic materials \cite{Stratton1941,BH1983} is straightforward
with the vector Helmholtz equation requiring only a simple redefinition: $\kappa = \ko\sqrt{\epsr\mur}$, where $\mur$ is
the relative permeability scalar, also frequency dependent. Extension to homogeneous bi-isotropic materials, which require the solution
of
\begin{equation}
\label{Eq3}
\nabla\times\left[\nabla\times{\#F}(\#r)\right]+i\ko(\zeta-\xi)\nabla\times{\#F}(\#r)-\ko^2(\epsr\mur-\zeta\xi){\#F}(\#r)=\#0\,
\end{equation}
with $\xi$ and $\zeta$ as frequency-dependent magnetoelectric scalars,
came much later \cite{Bohren1974}; even so, the solutions of Eq.~\r{Eq3} are simple combinations
of the solutions of Eq.~\r{Eq2} \cite{Bohren1974,Beltramibook}.

Vector spherical wavefunctions for homogeneous anisotropic materials have not been found in closed form.
A synthetic approach that requires finding vector cartesian wavefunctions \cite{EAB}
and subsequent transformation to spherical coordinates \cite{BKS} does not yield closed-form expressions for the
vector spherical wavefunctions \cite{Ren1993,GWLG}. This is because a spherical coordinate system
is generally incompatible with the cartesian nature of anisotropy.

An exception is furnished by the Minkowski constitutive relations of a simply moving material that is an isotropic dielectric-magnetic
material in its co-moving frame of reference. These constitutive relations can be stated as
\begin{eqnarray}
&&\#D\ro = \epso\epsr\left[\gamma\=I+(1-\gamma)\hat{\#v}\hat{\#v}\right]\.\#E\ro +\beta\hat{\#v}\times\#H\ro\,\\
&&\#B\ro = \muo\mur\left[\gamma\=I+(1-\gamma)\hat{\#v}\hat{\#v}\right]\.\#H\ro -\beta\hat{\#v}\times\#E\ro\,
\end{eqnarray}
where $\epso$ and $\muo$ are, respectively, the permittivity and the permeability
of free space;  $\epsr$, $\mur$, $\gamma$, and $\beta$ are real-valued scalars; the unit vector $\hat{\#v}$ is parallel to the velocity of the
material; and $\=I$ is the identity dyadic. By a  combination of two transformations \cite{LMcpl}, the frequency-domain Maxwell equations
in this material can be converted to frequency-domain Maxwell equations in a homogeneous, isotropic, dielectric-magnetic
material \cite{Danov2011}. The same algebraic manipulation is possible for a material \red{whose
constitutive relations are inspired by those of} gravitationally affected vacuum and
which can   potentially be fabricated  by properly dispersing electrically small bent-wire and other
complex inclusions of different shapes and materials
in some host materials \cite{LMcpl}.

We show in this paper that a similar combination of  transformations can be fruitfully applied to the more general material
described by the constitutive relations
\begin{eqnarray}
\label{con-D}
&&\#D\ro=\epso\epsr \,\=S\.\=A \.\=A \.\=S^{-1}\.\#E\ro
-{\co^{-1}}\,\#\Gamma \times\#H\ro\,,\\
&&\#B\ro=\muo\mur \,\=S\.\=A \.\=A \.\=S^{-1}\.\#H\ro
+{\co^{-1}}\,\#\Gamma \times\#E\ro\,.
\label{con-B}
\end{eqnarray}
In these equations,  $c_0=1/\sqrt{\epso\muo}$ is the speed of light in free space;
$\epsr \ne0$ and $\mur \ne0$ are complex-valued scalars; $\=S$ is either a rotation dyadic
or a product of rotation dyadics; the diagonal dyadic \red{$\=A$
is} positive definite with real-valued elements; and the gyrotropic vector $\#\Gamma$ can have complex-valued components.
For later use, we define the free-space impedance $\etao=\sqrt{\muo/\epso}$ and the relative
impedance $\eta_r(\omega)=\sqrt{\mu_r(\omega)/\eps_r(\omega)}$. The dependences of the
constitutive variables $\epsr$, $\mur$, $\=A$, and $\#\Gamma$ on the angular frequency $\omega>0$ have not been mentioned
for notational compactness, and an $\exp(-i\omega t)$
dependence on time $t$ is implicit.

Accordingly, using a set of three transformations developed
in Sec.~\ref{Transformations},
we determine closed-form expressions for
vector spherical wavefunctions applicable to a material with the
constitutive relations \r{con-D} and \r{con-B}, as shown in Sec.~\ref{VSWF}.
The derived vector spherical wavefunctions are used in Sec.~\ref{Tmatrix} to formulate
the T matrix for scattering by a three-dimensional object
composed of the chosen material. Section~\ref{radfields} is devoted to the radiation field of
an electric source current density distribution in the embedded material: a closed-form coordinate-free
expression of the dyadic Green function is obtained, followed by the elucidation of its singularity
behavior and its representation in a bilinear form involving the derived vector spherical wavefunctions.

\section{Transformations}\label{Transformations}
Representing the position vector  $\#r\equiv(x,y,z)\equiv(r,\theta,\phi)$, and denoting the source current density by $\#J\ro$,
we aim to solve
\begin{eqnarray}
\label{Faraday1}
&&\curl\#E\ro=i\omega\#B\ro\,,\\
\label{Ampere1}
&& \curl\#H\ro=-i\omega\#D\ro+\#J\ro\,,
\end{eqnarray}
when Eqs.~\r{con-D} and \r{con-B} have been substituted therein.
This is done after applying three transformations in succession so that we have to deal with effectively a homogeneous, isotropic,
dielectric-magnetic
material.

\subsection{First transformation} \l{1st_transformation}
Defining the auxiliary  fields \cite{LWmotl}
\begin{eqnarray}
\label{ero}
&&\#e\ro=\#E\ro\,\exp\le-i\ko\,\#\Gamma \.\#r\ri\,,\\
\label{hro}
&&\#h\ro=\#H\ro\,\exp\le-i\ko\,\#\Gamma \.\#r\ri\,,
\end{eqnarray}
and auxiliary source current density
\begin{equation}
\label{jro}
\#j\ro=\#J\ro\,\exp\le-i\ko\,\#\Gamma \.\#r\ri\,,
\end{equation}
and noting that $\ko=\omega/\co$, we transform Eqs.~\r{Faraday1} and \r{Ampere1} to
\begin{eqnarray}
\label{Faraday2}
&&\curl\#e\ro=i\omega\muo\mur \,\=S\.\=A \.\=A \.\=S^{-1}\.\#h\ro\,,\\
\label{Ampere2}
&& \curl\#h\ro=-i\omega\epso\epsr \,\=S\.\=A \.\=A \.\=S^{-1}\.\#e\ro+\#j\ro\,.
\end{eqnarray}

\subsection{Second transformation} \l{2nd_transformation}
Next, we define another set of auxiliary quantities
\begin{eqnarray}
\label{erpo}
&&\#e^\prime\rpo=\=S^{-1}\.\#e\ro\,,\\
\label{hrpo}
&&\#h^\prime\rpo=\=S^{-1}\.\#h\ro\,,\\
\label{jrpo}
&&\#j^\prime\rpo=\=S^{-1}\.\#j\ro\,,
\end{eqnarray}
and coordinates $\#r^\prime\equiv(x^\prime,y^\prime,z^\prime)\equiv(r^\prime,\theta^\prime,\phi^\prime)$ such that
\begin{equation}
\#r^\prime=\=S\.\#r
\end{equation}
in order to transform Eqs.~\r{Faraday2} and \r{Ampere2} to
\begin{eqnarray}
\label{Faraday3}
&&\curlp\#e^\prime\rpo=i\omega\muo\mur \,\=A \.\=A \. \#h^\prime\rpo\,,\\
\label{Ampere3}
&&\curlp\#h^\prime\rpo=-i\omega\epso\epsr \,\=A \.\=A \. \#e^\prime\rpo+\#j^\prime\rpo\,,
\end{eqnarray}
because
\begin{equation}
\=S^{-1}\.\left(\curl\=I\right)\.\=S=\curlp\=I\,.
\end{equation}
This transformation merely amounts to a rigid rotation of the coordinate system, and can be avoided by an
appropriate choice of the $(x,y,z)$ coordinate system.

\subsection{Third transformation} \l{3rd_transformation}
Finally, we make use of an affine transformation \cite{Gray,LWijaem}
associated with the scaling of space as
\begin{equation}
\~r=\=A \.\#r^\prime\,,
\end{equation}
where $\~r\equiv(\tilde{x},\tilde{y},\tilde{z})\equiv(\tilde{r},\tilde{\theta},\tilde{\phi})$,
and define the final set of auxiliary quantities as
\begin{eqnarray}
\label{erto}
&& \~e\rpo=\=A \.\#e^\prime\rto\,,\\
\label{hrto}
&&\~h\rpo=\=A \.\#h^\prime\rto\,,\\
\label{jrto}
&&\~j\rpo=\le\adj\=A\ri \.\#j^\prime\rto\,,
\end{eqnarray}
to obtain
\begin{eqnarray}
\label{Faraday4}
&&\curlp\~e\rpo=i\omega\muo\mur \,A \,\~h\rpo\,,\\
\label{Ampere4}
&&\curlp\~h\rpo=-i\omega\epso\epsr \, A \, \~e\rpo+\~j\rpo\,,
\end{eqnarray}
where $A $ is the determinant of $\=A $. Equations \r{Faraday4}
and \r{Ampere4} are identical to analogous equations for a homogeneous, isotropic,
dielectric-magnetic material.

From these two equations, we
get
\red{
\begin{eqnarray}
\label{Helmholtz4e}
&&\curlp\les\curlp\~e\rpo\ris-\ko^2 \epsr \mur \,A^2 \, \~e\rpo =i\omega\muo\mur\,A\,\~j\rpo
\,,\\
\label{Helmholtz4h}
&&\curlp\les\curlp\~h\rpo\ris-\ko^2 \epsr \mur \,A^2 \, \~h\rpo =\curlp\~j\rpo
\,.
\end{eqnarray}
}
In the absence of a source current density, these equations are the same as
Eq.~\r{Eq1}.

\section{Field representation in a source-free region} \label{VSWF}
Homogeneous solutions of the Eqs.~\r{Helmholtz4e} and  \r{Helmholtz4h}  in the spherical coordinate system were obtained by Clebsch \cite{Logan1965}.
Thus, in the absence of the source current density,
\begin{equation}
\label{etilde}
\~e\rpo=\sum_{j=1,3}\sum_{s=e,o}\sum_{n=0}^{\infty}\sum_{m=0}^{n}\les
a_{smn}^{(j)} \,\#L_{smn}^{(j)}\le{k\#r^\prime}\ri
+
b_{smn}^{(j)} \,\#M_{smn}^{(j)}\le{k\#r^\prime}\ri
+
c_{smn}^{(j)} \,\#N_{smn}^{(j)}\le{k\#r^\prime}\ri\ris\,
\end{equation}
and
\begin{equation}
\label{htilde}
\~h\rpo=\sum_{j=1,3}\sum_{s=e,o}\sum_{n=0}^{\infty}\sum_{m=0}^{n}\les
\alpha_{smn}^{(j)} \,\#L_{smn}^{(j)}\le{k\#r^\prime}\ri
+
\beta_{smn}^{(j)} \,\#M_{smn}^{(j)}\le{k\#r^\prime}\ri
+
\gamma_{smn}^{(j)} \,\#N_{smn}^{(j)}\le{k\#r^\prime}\ri\ris\,,
\end{equation}
where the wavenumber
\begin{equation}
k=\ko {A }\sqrt{\epsr}\sqrt{ \mur}\,;
\end{equation}
the standard vector spherical wave functions \cite{MF}
\begin{eqnarray}
\label{Ldef1}
&&\#L_{smn}^{(1)}\le{k\#r^\prime}\ri=\frac{1}{k} \nabla^\prime\les{Y_{smn}({\theta^\prime}, {\phi^\prime})\,j_n(kr^\prime)}\ris\,,
\\
\label{Ldef3}
&&\#L_{smn}^{(3)}\le{k\#r^\prime}\ri=\frac{1}{k} \nabla^\prime\les{Y_{smn}({\theta^\prime}, {\phi^\prime})\,h_n^{(1)}(kr^\prime)}\ris\,,
\\
\label{Mdef1}
&&\#M_{smn}^{(1)}\le{k\#r^\prime}\ri=\curlp \les\#r^\prime\,
{Y_{smn}({\theta^\prime}, {\phi^\prime})\,j_n(kr^\prime)}\ris\,,
\\
\label{Mdef3}
&&\#M_{smn}^{(3)}\le{k\#r^\prime}\ri=\curlp \les\#r^\prime\,
{Y_{smn}({\theta^\prime}, {\phi^\prime})\,h_n^{(1)}(kr^\prime)}\ris\,,
\\
\label{Ndef}
&&\#N_{smn}^{(j)}\le{k\#r^\prime}\ri=\frac{1}{k}\curlp\#M_{smn}^{(j)}\le{k\#r^\prime}\ri\,;
\end{eqnarray}
the tesseral harmonics
\begin{eqnarray}
\label{Ydefe}
&&Y_{emn}({\theta^\prime}, {\phi^\prime})= P_n^m\le{\cos\theta^\prime}\ri\,\cos\le{m\phi^\prime}\ri\,,
\\
\label{Ydefo}
&&Y_{omn}({\theta^\prime}, {\phi^\prime})= P_n^m\le{\cos\theta^\prime}\ri\,\sin\le{m\phi^\prime}\ri\,;
\end{eqnarray}
$j_n(\.)$ is the spherical Bessel function of order $n$, $h_n^{(1)}(\.)$ is the spherical Hankel function of
the first kind and order $n$,
$P_n^m(\.)$ is the associated Legendre function
of order $n$ and degree $m$; and the coefficients $a_{smn}^{(j)}$, $b_{smn}^{(j)}$,  $c_{smn}^{(j)}$,
$\alpha_{smn}^{(j)}$, $\beta_{smn}^{(j)}$, and  $\gamma_{smn}^{(j)}$
have to be determined from various stipulations on the spatial variations of $\~e\rpo$ and $\~h\rpo$.

As the identities $\curlp{\#L_{smn}^{(j)}\le{k\#r^\prime}\ri}\equiv\#0$,
$\divp{\#M_{smn}^{(j)}\le{k\#r^\prime}\ri}\equiv0$, and $\divp{\#N_{smn}^{(j)}\le{k\#r^\prime}\ri}\equiv0$
 follow from the definitions \r{Ldef1}--\r{Ndef},  substitution of
 Eqs.~\r{etilde} and \r{htilde} in Eqs.~\r{Faraday4} and \r{Ampere4}
yields the simplifications
\begin{equation}
a_{smn}^{(j)}\equiv0\,,\quad \alpha_{smn}^{(j)}\equiv0\,.
\end{equation}
Furthermore, as
$\curlp\#M_{smn}^{(j)}\le{k\#r^\prime}\ri=k\#N_{smn}^{(j)}\le{k\#r^\prime}\ri$ and
$\curlp\#N_{smn}^{(j)}\le{k\#r^\prime}\ri=k\#M_{smn}^{(j)}\le{k\#r^\prime}\ri$, the same
substitution also yields the simplifications
\begin{equation}
\beta_{smn}^{(j)}=-\frac{ic_{smn}^{(j)}}{\etao\eta_r(\omega)}\,,\quad
\gamma_{smn}^{(j)}=-\frac{ib_{smn}^{(j)}}{\etao\eta_r(\omega)}\,.
\end{equation}
Accordingly,  Eqs.~\r{etilde} and \r{htilde} can be rewritten as follows:
\begin{eqnarray}
\label{etilde-final}
&&\~e\rpo=\sum_{j=1,3}\sum_{s=e,o}\sum_{n=1}^{\infty}\sum_{m=0}^{n}\les
b_{smn}^{(j)} \,\#M_{smn}^{(j)}\le{k\#r^\prime}\ri
+
c_{smn}^{(j)} \,\#N_{smn}^{(j)}\le{k\#r^\prime}\ri\ris\,,
\\
\label{htilde-final}
&&\~h\rpo=-\frac{i}{\etao\eta_r(\omega)}
\sum_{j=1,3}\sum_{s=e,o}\sum_{n=1}^{\infty}\sum_{m=0}^{n}\les
c_{smn}^{(j)} \,\#M_{smn}^{(j)}\le{k\#r^\prime}\ri
+
b_{smn}^{(j)} \,\#N_{smn}^{(j)}\le{k\#r^\prime}\ri\ris\,.
\end{eqnarray}
Let us note here that the set of functions $\#M_{smn}^{(j)}\le{k\#r^\prime}\ri$ and $\#N_{smn}^{(j)}\le{k\#r^\prime}\ri$
is complete \cite{AH1986}.

\subsection{Vector spherical wavefunctions}
Inverting the relations \r{erto} and \r{hrto}, we get
\begin{eqnarray}
\label{erp-final}
&&\#e^\prime\rpo= \=A^{-1}\.
\sum_{j=1,3}\sum_{s=e,o}\sum_{n=1}^{\infty}\sum_{m=0}^{n}\les
b_{smn}^{(j)} \,\#M_{smn}^{(j)}\le{k\=A^{-1}\.\#r^\prime}\ri
+
c_{smn}^{(j)} \,\#N_{smn}^{(j)}\le{k\=A^{-1}\.\#r^\prime}\ri\ris\,,
\\
\nonumber
&&\#h^\prime\rpo= -\frac{i}{\etao\eta_r}\,\=A^{-1}\.
\sum_{j=1,3}\sum_{s=e,o}\sum_{n=1}^{\infty}\sum_{m=0}^{n}\les
c_{smn}^{(j)} \,\#M_{smn}^{(j)}\le{k\=A^{-1}\.\#r^\prime}\ri\right.
\\
\label{hrp-final}
&&\qquad\qquad\qquad\qquad\qquad\qquad\qquad\qquad\qquad\qquad\qquad
\left.+
b_{smn}^{(j)} \,\#N_{smn}^{(j)}\le{k\=A^{-1}\.\#r^\prime}\ri\ris\,
\end{eqnarray}
from Eqs.~\r{etilde-final} and \r{htilde-final}.
Next, an inversion involving Eqs.~\r{erpo} and \r{hrpo}  yields
 \begin{eqnarray}
\nonumber
&&\#e\ro=\=S\. \=A^{-1}\.
\sum_{j=1,3}\sum_{s=e,o}\sum_{n=1}^{\infty}\sum_{m=0}^{n}\les
b_{smn}^{(j)} \,\#M_{smn}^{(j)}\le{k\=A^{-1}\.\=S\.\#r}\ri\right.
\\
\label{ero-final}
&&\qquad\qquad\qquad\qquad\qquad\qquad\qquad\qquad\qquad\qquad\qquad
\left.+
c_{smn}^{(j)} \,\#N_{smn}^{(j)}\le{k\=A^{-1}\.\=S\.\#r }\ri\ris\,,
\\
\nonumber
&&\#h\ro= -\frac{i}{\etao\eta_r}\,\=S\.\=A^{-1}\.
\sum_{j=1,3}\sum_{s=e,o}\sum_{n=1}^{\infty}\sum_{m=0}^{n}\les
c_{smn}^{(j)} \,\#M_{smn}^{(j)}\le{k\=A^{-1}\.\=S\.\#r }\ri\right.
\\
\label{hro-final}
&&\qquad\qquad\qquad\qquad\qquad\qquad\qquad\qquad\qquad\qquad\qquad
\left.+
b_{smn}^{(j)} \,\#N_{smn}^{(j)}\le{k\=A^{-1}\.\=S\.\#r }\ri\ris\,,
\end{eqnarray}
from where we finally obtain
\begin{eqnarray}
\nonumber
&&\#E\ro=\=S\. \=A^{-1}\.
\sum_{j=1,3}\sum_{s=e,o}\sum_{n=1}^{\infty}\sum_{m=0}^{n}\les
b_{smn}^{(j)} \,\#M_{smn}^{(j)}\le{k\=A^{-1}\.\=S\.\#r}\ri\right.
\\
\label{Ero-fina}
&&\qquad\qquad\qquad\qquad\qquad\qquad\qquad
\left.+
c_{smn}^{(j)} \,\#N_{smn}^{(j)}\le{k\=A^{-1}\.\=S\.\#r }\ri\ris\exp\le{i\ko\,\#\Gamma\.\#r}\ri\,,
\\
\nonumber
&&\#H\ro= -\frac{i}{\etao\eta_r}\,\=S\.\=A^{-1}\.
\sum_{j=1,3}\sum_{s=e,o}\sum_{n=1}^{\infty}\sum_{m=0}^{n}\les
c_{smn}^{(j)} \,\#M_{smn}^{(j)}\le{k\=A^{-1}\.\=S\.\#r }\ri\right.
\\
\label{Hro-fina}
&&\qquad\qquad\qquad\qquad\qquad\qquad\qquad
\left.+
b_{smn}^{(j)} \,\#N_{smn}^{(j)}\le{k\=A^{-1}\.\=S\.\#r }\ri\ris\exp\le{i\ko\,\#\Gamma\.\#r}\ri\,.
\end{eqnarray}

Thus, for the material described by Eqs.~\r{con-D} and \r{con-B},
 the desired vector spherical wavefunctions in the spherical coordinates $(r,\theta,\phi)$ are
\begin{equation}
\label{mdefj}
\#m_{smn}^{(j)}(\#r)= \exp\le{i\ko\,\#\Gamma\.\#r}\ri
\=S\. \=A^{-1}\.\#M_{smn}^{(j)}\le{k\=A^{-1}\.\=S\.\#r}\ri
\end{equation}
and
\begin{equation}
\label{ndefj}
\#n_{smn}^{(j)}(\#r)=\exp\le{i\ko\,\#\Gamma\.\#r}\ri
\=S\. \=A^{-1}\.\#N_{smn}^{(j)}\le{k\=A^{-1}\.\=S\.\#r}\ri\,.
\end{equation}
Equations \r{Ero-fina} and \r{Hro-fina} can then be compactly written as
\begin{eqnarray}
\label{Ero-final}
&&\#E\ro=
\sum_{j=1,3}\sum_{s=e,o}\sum_{n=1}^{\infty}\sum_{m=0}^{n}\les
b_{smn}^{(j)} \,\#m_{smn}^{(j)}\le{\#r}\ri
+
c_{smn}^{(j)} \,\#n_{smn}^{(j)}\le{\#r }\ri\ris\,,
\\
\label{Hro-final}
&&\#H\ro= -\frac{i}{\etao\eta_r}\,
\sum_{j=1,3}\sum_{s=e,o}\sum_{n=1}^{\infty}\sum_{m=0}^{n}\les
c_{smn}^{(j)} \,\#m_{smn}^{(j)}\le{\#r }\ri\
+
b_{smn}^{(j)} \,\#n_{smn}^{(j)}\le{\#r }\ri\ris\,.
\end{eqnarray}
The completeness of the set of functions $\#m_{smn}^{(j)}\le{\#r}\ri$ and $\#n_{smn}^{(j)}\le{\#r}\ri$ follows from the
completeness of the set of functions
$\#M_{smn}^{(j)}\le{k\#r^\prime}\ri$ and $\#N_{smn}^{(j)}\le{k\#r^\prime}\ri$
\cite{AH1986}.

\subsection{Closed-form expressions}
Without any loss of generality, we set
\begin{equation}
\left.\begin{array}{l}
\=S=\=I\\
\=A=\alpha_x^{-1}\ux\ux+\alpha_y^{-1}\uy\uy+\uz\uz
\end{array}\right\}\,,
\label{stillgen}
\end{equation}
where $\alpha_x>0$ and $\alpha_y>0$. In order to avoid extremely cumbersome expressions
for the vector spherical wavefunctions $\#m_{smn}^{(j)}\le{\#r}\ri$ and
$\#n_{smn}^{(j)}\le{\#r}\ri$, we also define wavefunctions $\newM_{smn}^{(j)}\le{\#r}\ri$ and
$\newN_{smn}^{(j)}\le{\#r}\ri$ via
\begin{eqnarray}
\label{newMdefj}
&&\#m_{smn}^{(j)}\le{\#r}\ri= \=A^{-1}\.\newM_{smn}^{(j)}\le{\#r}\ri= \=B\.\newM_{smn}^{(j)}\le{\#r}\ri
\,,
\\
\label{newNdefj}
&&\#n_{smn}^{(j)}\le{\#r}\ri= \=A^{-1}\.\newN_{smn}^{(j)}\le{\#r}\ri= \=B\.\newN_{smn}^{(j)}\le{\#r}\ri\,,
\end{eqnarray}
with
\begin{equation}
\=B=\alpha_x \ux\ux+\alpha_y \uy\uy+\uz\uz\,.
\end{equation}

Using Eqs.~\r{Mdef1}--\r{Ydefo}, \r{mdefj}, and \r{ndefj}, we obtain
\begin{eqnarray}
\nonumber
&&\newM_{smn}^{(j)}(\#r)=\exp\les{i\ko\,\#\Gamma\.\#r}\ris
\frac{\calJ_n^{(j)}(k\#r)}{f_1(\phi)}
\\
&&\nonumber\quad\times
\lec
\hat{\#r} \left[
\frac{f_4(\phi)-f_1^2(\phi)}{f_2(\theta,\phi)}\st\ct  \,\calQ_{smn}(\theta,\phi)
-(\alpha_x-\alpha_y)\st\sp\cp \,\calR_{smn}(\theta,\phi)\right]
\right.
\\
&&\nonumber\qquad
+
\utheta \left[
\frac{f_4(\phi)\cos^2\theta+f_1^2(\phi)\sin^2\theta}{f_2(\theta,\phi)} \,\calQ_{smn}(\theta,\phi)
-(\alpha_x-\alpha_y)\ct\sp\cp\,\calR_{smn}(\theta,\phi)\right]
\\
&&\qquad
+
\left.
\uphi \left[
-\,\frac{\alpha_x-\alpha_y}{f_2(\theta,\phi)}\ct\sp\cp \,\calQ_{smn}(\theta,\phi)
- f_4(\phi)  \,\calR_{smn}(\theta,\phi)\right]
\ric
\,
\end{eqnarray}
and
\begin{eqnarray}
\nonumber
&&\newN_{smn}^{(j)}(\#r)=\exp\les{i\ko\,\#\Gamma\.\#r}\ris
\\
&&\quad
\nonumber
\times\Bigg(
\hat{\#r}\lec \frac{\calJ_n^{(j)}(k\#r)}{kr }\,
\les\frac{\cos^2\theta+f_4(\phi)\sin^2\theta}{f_2^2(\theta,\phi)}\ris \,
\calP_{smn}(\theta,\phi)\,
\right.\\ &&\nonumber\qquad\quad\left.
+\frac{\calK_n^{(j)}(k\#r)}{f_1(\phi)}
\left[
\frac{f_4(\phi)-f_1^2(\phi)}{f_2(\theta,\phi)}\st\ct \,\calR_{smn}(\theta,\phi)  +
(\alpha_x-\alpha_y)\st\sp\cp\,\calQ_{smn}(\theta,\phi)\right]\ric
\\
&&
\nonumber\qquad
+\utheta\lec\frac{\calJ_n^{(j)}(k\#r)}{kr}\,
\les\frac{f_4(\phi)-1}{ f_2^2(\theta,\phi)}\st\ct\ris\,
\calP_{smn}(\theta,\phi)\,
\right.\\ &&\nonumber\qquad\quad\left.
+\frac{\calK_n^{(j)}(k\#r)}{f_1(\phi)}
\left[
\frac{f_4(\phi)\cos^2\theta+f_1^2(\phi)\sin^2\theta}{f_2(\theta,\phi)}\,
\calR_{smn}(\theta,\phi)  +
(\alpha_x-\alpha_y)\ct\sp\cp\,\calQ_{smn}(\theta,\phi)\right]\ric
\\
&&\nonumber\qquad
+\uphi\lec
-\,\frac{\calJ_n^{(j)}(k\#r)}{kr}\,
\les\frac{\alpha_x-\alpha_y}{f_2^2(\theta,\phi)}\st\sp\cp\ris\,
\calP_{smn}(\theta,\phi)\,
\right.\\ && \qquad\quad\left.
+\frac{\calK_n^{(j)}(k\#r)}{f_1(\phi)}
\left[
-\,\frac{\alpha_x-\alpha_y}{f_2(\theta,\phi)}\ct\sp\cp\,\calR_{smn}(\theta,\phi)
+f_4(\phi)\,\calQ_{smn}(\theta,\phi)\right]\ric\Bigg)\,.
\end{eqnarray}
Furthermore,
\begin{eqnarray}
\nonumber
&&
\=B=\les{f_4(\phi)\sin^2\theta+\cos^2\theta}\ris\ur\ur
+\les{f_4(\phi)\cos^2\theta+\sin^2\theta}\ris \utheta\utheta
+\les{\alpha_x+\alpha_y-f_4(\phi)}\ris\uphi\uphi
\\
\nonumber
&&\qquad+\les{f_4(\phi)-1}\ris\st\ct\le\ur\utheta+\utheta\ur\ri
\\
\label{Bdef}
&&\qquad -\le\alpha_x-\alpha_y\ri\sp\cp\les\st\le\ur\uphi+\uphi\ur\ri  +\ct\le\utheta\uphi+\uphi\utheta\ri\ris
\,.
\end{eqnarray}

In the foregoing expressions, the following angular functions have been used:
\begin{eqnarray}
&&
f_1(\phi) =+\left(\alpha_x^2\cos^2\phi+\alpha_y^2\sin^2\phi\right)^{1/2}\,,
\\
&&
f_2(\theta,\phi) =+\left[f_1^2(\phi)\sin^2\theta+\cos^2\theta\right]^{1/2}\,,
\\
&&
f_3(\phi)=\tan^{-1}\left(\frac{\alpha_y}{\alpha_x}\tan\phi\right)\,,
\\
&&
f_4(\phi)=\alpha_x \cos^2\phi+\alpha_y \sin^2\phi\,.
\end{eqnarray}
The angle $f_3(\phi)$ must lie in the same quadrant as its argument. Additionally,
\begin{eqnarray}
&&
\calJ_n^{(1)}(k\#r)=j_n\left[krf_2(\theta,\phi)\right]\,,
\\
&&
\calJ_n^{(3)}(k\#r)=h_n^{(1)}\left[krf_2(\theta,\phi)\right]\,,
\\
&&
\calK_n^{(j)}(k\#r)=\frac{n+1}{krf_2(\theta,\phi)}\, \calJ_n^{(j)}(k\#r) -\calJ_{n+1}^{(j)}(k\#r)\,,\quad
j\in\lec1,3\ric\,,
\\
&&
\calP_{smn}(\theta,\phi)=n(n+1)P_n^m\left[\frac{\cos\theta}{f_2(\theta,\phi)}\right]\calV_{sm}(\phi)\,,
\\
&&
\calQ_{smn}(\theta,\phi)=mP_n^m\left[\frac{\cos\theta}{f_2(\theta,\phi)}\right]\frac{f_2(\theta,\phi)}{f_1(\phi)\sin\theta}\,
\calU_{sm}(\phi)\,,
\\
&&
\nonumber
\calR_{smn}(\theta,\phi)=\frac{1}{f_1(\phi)\sin\theta}\left\{
(n-m+1) {f_2(\theta,\phi)} P_{n+1}^m\left[\frac{\cos\theta}{f_2(\theta,\phi)}\right]
\right.
\\
&&
\qquad
\left.
-
(n+1) {\cos\theta} {P_n^m}\left[\frac{\cos\theta}{f_2(\theta,\phi)}\right]\right\}\calV_{sm}(\phi)\,,
\\
&&
\calU_{sm}(\phi)=\lec\begin{array}{c}-\sin \left[mf_3(\phi)\right]\\\cos\left[mf_3(\phi)\right]\end{array}\ric
\,,
\quad s=\left\{\begin{array}{c}e\\{o}\end{array}\right.\,,
\\
&&
\calV_{sm}(\phi)=\lec\begin{array}{c}\cos \left[mf_3(\phi)\right]\\\sin\left[mf_3(\phi)\right]\end{array}\ric
\,,
\quad s=\left\{\begin{array}{c}e\\{o}\end{array}\right.\,.
\end{eqnarray}

\subsubsection{Special case: $\alpha_x=\alpha_y=\alpha$}
When the  diagonal dyadic $\=A$ is simplified from  biaxial ($\alpha_x\ne\alpha_y$) to
uniaxial ($\alpha_x=\alpha_y=\alpha$), the functions $f_1$ to $f_4$ simplify considerably.
The wavefunctions $\newM_{smn}^{(j)}(\#r)$ and $\newN_{smn}^{(j)}(\#r)$ then simplify as follows:
\begin{eqnarray}
\nonumber
&&\newM_{smn}^{(j)}(\#r)=\exp\les{i\ko\,\#\Gamma\.\#r}\ris
\tcalJ_n^{(j)}(k\#r)
\lec
-\,\hat{\#r} \frac{\alpha-1}{f_5(\theta)}\st\ct  \,\tcalQ_{smn}(\theta,\phi)
\right.
\\
&&\qquad
+
\utheta
\frac{ \alpha\sin^2\theta+\cos^2\theta}{f_5(\theta)} \,\tcalQ_{smn}(\theta,\phi)
-
\left.
\uphi  \,\tcalR_{smn}(\theta,\phi)
\ric
\,,
\end{eqnarray}
\begin{eqnarray}
\nonumber
&&\newN_{smn}^{(j)}(\#r)=\exp\les{i\ko\,\#\Gamma\.\#r}\ris
\\
&&\quad
\nonumber
\times\Bigg\{
\hat{\#r}\les
\frac{\tcalJ_n^{(j)}(k\#r)}{kr }\,
 \frac{f_6(\theta)}{f_5^2(\theta)}  \,
\tcalP_{smn}(\theta,\phi)
-\, \tcalK_n^{(j)}(k\#r)
\frac{\alpha-1}{f_5(\theta)}\st\ct \,
\tcalR_{smn}(\theta,\phi)  \ris
\\
&&
\nonumber\qquad
+\utheta\les
\frac{\tcalJ_n^{(j)}(k\#r)}{kr}\,
 \frac{\alpha-1}{ f_5^2(\theta)}\st\ct \,
\tcalP_{smn}(\theta,\phi)
+ \tcalK_n^{(j)}(k\#r)
\frac{f_6(\theta)}{f_5(\theta)}\,
\tcalR_{smn}(\theta,\phi)   \ris
\\
&&\qquad
+\uphi \,
\tcalK_n^{(j)}(k\#r) \,\tcalQ_{smn}(\theta,\phi) \Bigg\}\,.
\end{eqnarray}
Similarly, $\=B$ also simplifies to
\begin{eqnarray}
&&
\=B={f_6(\theta)}\,\ur\ur
+\les{\alpha+1-f_6(\theta)}\ris\utheta\utheta
+\alpha\, \uphi\uphi
+(\alpha-1)\st\ct
\le\ur\utheta+\utheta\ur\ri\,.
\end{eqnarray}

The following shorthand notation has been used:
\begin{eqnarray}
&&
f_5(\theta)=+\left(\alpha^2\sin^2\theta+\cos^2\theta\right)^{1/2} \,,
\\
&&
f_6(\theta)=\alpha \sin^2\theta+\cos^2\theta\,,
\\
&&
\tcalJ_n^{(1)}(k\#r)=j_n\left[krf_5(\theta)\right]\,,
\\
&&
\tcalJ_n^{(3)}(k\#r)=h_n^{(1)}\left[krf_5(\theta)\right]\,,
\\
&&
\tcalK_n^{(j)}(k\#r)=\frac{n+1}{krf_5(\theta)}\,  \tcalJ_n^{(j)}(k\#r)
-\tcalJ_{n+1}^{(j)}(k\#r)\,,\quad
j\in\lec1,3\ric\,,
\\
&&
\tcalP_{smn}(\theta,\phi)=n(n+1)P_n^m\left[\frac{\cos\theta}{f_5(\theta)}\right]
\lec\begin{array}{c}\cos  \left(m \phi\right)\\\sin \left(m \phi\right)\end{array}\ric
\,,
\quad s=\left\{\begin{array}{c}e\\{o}\end{array}\right.\,,
\\
&&
\tcalQ_{smn}(\theta,\phi)=mP_n^m\left[\frac{\cos\theta}{f_5(\theta)}\right]\frac{f_5(\theta)}{\alpha\sin\theta}\,
\lec\begin{array}{c}-\sin \left(m \phi\right)\\\cos \left(m \phi\right)\end{array}\ric
\,,
\quad s=\left\{\begin{array}{c}e\\{o}\end{array}\right.\,,
\\
&&
\nonumber
\tcalR_{smn}(\theta,\phi)=\frac{1}{\alpha\sin\theta}\left\{
(n-m+1) {f_5(\theta)} P_{n+1}^m\left[\frac{\cos\theta}{f_5(\theta)}\right]
\right.
\\
&&
\qquad
\left.
-
(n+1) {\cos\theta} {P_n^m}\left[\frac{\cos\theta}{f_5(\theta)}\right]\right\}
\lec\begin{array}{c}\cos  \left(m \phi\right)\\\sin \left(m \phi\right)\end{array}\ric
\,,
\quad s=\left\{\begin{array}{c}e\\{o}\end{array}\right.\,.
\end{eqnarray}

\subsubsection{Special case: $\alpha_x=\alpha_y=1$}
Further simplification to $\=A = \=I$ leads to $\=B=\=I$, so that
\begin{eqnarray}
&&\#m_{smn}^{(j)}(\#r)=\newM_{smn}^{(j)}(\#r)=\exp\le{i\ko\,\#\Gamma\.\#r}\ri
\#M_{smn}^{(j)}(k\#r)\,,
\\
&&\#n_{smn}^{(j)}(\#r)=\newN_{smn}^{(j)}(\#r)=\exp\le{i\ko\,\#\Gamma\.\#r}\ri
\#N_{smn}^{(j)}(k\#r)\,.
\end{eqnarray}

\subsection{T matrix for scattering by a 3D anisotropic object}\label{Tmatrix}
For an example of the use of the derived vector spherical wavefunctions, let
 us now consider scattering by a 3D object that occupies the region $\sfV$ and has a surface $\sfS$. The material
occupying $\sfV$ obeys the constitutive relations \r{con-D} and \r{con-B}. The coordinate system is chosen
so that its origin lies inside $\sfV$ and the conditions \r{stillgen} hold. The external region is vacuous.

Suppose that an electromagnetic field is incident on the scatterer.  This incident field is purely arbitrary except that \red{its source must lie outside} 
a minimum sphere centered at the origin and circumscribing $\sfV$. It can then be represented
as \cite{BY1975}
\begin{eqnarray}
\label{Einc}
&&
\#E_{inc}\ro=\sum_{s=e,o}\sum_{n=1}^{\infty}\sum_{m=0}^{n}\,
D_{mn} \les A\smn^{(1)}\#M\smn^{(1)}(\ko\#r) + B\smn^{(1)}\#N\smn^{(1)}(\ko\#r)\ris\,,
\\
\label{Hinc}
&&
\#H_{inc}\ro=-\frac{i}{\etao}\sum_{s=e,o}\sum_{n=1}^{\infty}\sum_{m=0}^{n}\,
D_{mn} \les A\smn^{(1)}\#N\smn^{(1)}(\ko\#r) + B\smn^{(1)}\#M\smn^{(1)}(\ko\#r)\ris\,,
\end{eqnarray}
where the coefficients $A\smn^{(1)}$ and $B\smn^{(1)}$ are presumed to be known and
the normalization constant
\begin{equation}
D_{mn} = (2-\delta_{m0})\,\frac{2n+1}{4n(n+1)}\,\frac{(n-m)!}{(n+m)!}\,,
\end{equation}
involves the Kronecker delta $\delta_{pq}$. Outside the
circumscribing sphere, the scattered electromagnetic field is
represented by the sums
\begin{eqnarray}
&&
\label{Esca}
\#E_{sca}\ro=\sum_{s=e,o}\sum_{n=1}^{\infty}\sum_{m=0}^{n}\,
D_{mn} \les A\smn^{(3)}\#M\smn^{(3)}(\ko\#r) + B\smn^{(3)}\#N\smn^{(3)}(\ko\#r)\ris\,,
\\
\label{Hsca}
&&
\#H_{sca}\ro=-\frac{i}{\etao}\sum_{s=e,o}\sum_{n=1}^{\infty}\sum_{m=0}^{n}\,
D_{mn} \les A\smn^{(3)}\#N\smn^{(3)}(\ko\#r) + B\smn^{(3)}\#M\smn^{(3)}(\ko\#r)\ris\,,
\end{eqnarray}
where the unknown coefficients $A\smn^{(3)}$ and $B\smn^{(3)}$ have to be determined.
The underlying presumption here is that the source of the incident field lies so far away from the
scattering object that any action of the scattered field on that source is vanishingly small.

Application of the Ewald--Oseen extinction theorem followed by some analytic-continuation arguments
leads to the following four relationships \cite{BY1975,LVV1985}:
\begin{eqnarray}
&&\nonumber
A\smn^{(1)}= -\frac{i\ko^2}{\pi}
\int_\sfS d^2{\#r_p}\lec
\les\#n(\#r_p)\times \#E_{int}(\#r_p)\ris\.\#N\smn^{(3)}(\ko\#r_p)
\right.
\\
\label{Asmn1}
&&\qquad\left.
+i\etao \les\#n(\#r_p)\times \#H_{int}(\#r_p)\ris\.\#M\smn^{(3)}(\ko\#r_p)
\ric\,,
\\
&&\nonumber
B\smn^{(1)}= -\frac{i\ko^2}{\pi}
\int_\sfS d^2{\#r_p}\lec
\les\#n(\#r_p)\times \#E_{int}(\#r_p)\ris\.\#M\smn^{(3)}(\ko\#r_p)
\right.
\\
\label{Bsmn1}
&&\qquad\left.
+i\etao \les\#n(\#r_p)\times \#H_{int}(\#r_p)\ris\.\#N\smn^{(3)}(\ko\#r_p)
\ric\,,
\\
&&\nonumber
A\smn^{(3)}= \frac{i\ko^2}{\pi}
\int_\sfS d^2{\#r_p}\lec
\les\#n(\#r_p)\times \#E_{int}(\#r_p)\ris\.\#N\smn^{(1)}(\ko\#r_p)
\right.
\\
\label{Asmn3}
&&\qquad\left.
+i\etao \les\#n(\#r_p)\times \#H_{int}(\#r_p)\ris\.\#M\smn^{(1)}(\ko\#r_p)
\ric\,,
\\
&&\nonumber
B\smn^{(3)}= \frac{i\ko^2}{\pi}
\int_\sfS d^2{\#r_p}\lec
\les\#n(\#r_p)\times \#E_{int}(\#r_p)\ris\.\#M\smn^{(1)}(\ko\#r_p)
\right.
\\
\label{Bsmn3}
&&\qquad\left.
+i\etao \les\#n(\#r_p)\times \#H_{int}(\#r_p)\ris\.\#N\smn^{(1)}(\ko\#r_p)
\ric\,.
\end{eqnarray}
Here, $\#n(\#r_p)$ is the unit outward normal to $\sfS$ at $\#r_p\in\sfS$, whereas
$\#E_{int}(\#r)$ and $\#H_{int}(\#r)$ are the fields excited inside the scatterer.

A representation of the internal fields emerges from Eqs.~\r{Ero-final}, \r{Hro-final}, \r{newMdefj},
and \r{newNdefj}
as
\begin{eqnarray}
\label{Eint}
&&\#E_{int}\ro=   \=B\.
\sum_{s=e,o}\sum_{n=1}^{\infty}\sum_{m=0}^{n}\les
b_{smn}^{(1)} \,\newM_{smn}^{(1)}\le{\#r}\ri
+
c_{smn}^{(1)} \,\newN_{smn}^{(1)}\le{\#r }\ri\ris\,,
\\
\label{Hint}
&&\#H_{int}\ro= -\frac{i}{\etao\eta_r}\, \=B\.
\sum_{s=e,o}\sum_{n=1}^{\infty}\sum_{m=0}^{n}\les
c_{smn}^{(1)} \,\newM_{smn}^{(1)}\le{\#r }\ri +
b_{smn}^{(1)} \,\newN_{smn}^{(1)}\le{\#r }\ri\ris \,,
\end{eqnarray}
because the internal fields must be regular at the origin $\#0\in\sfV$.

Substitution of Eqs.~\r{Eint} and \r{Hint} in Eqs.~\r{Asmn1} and \r{Bsmn1} leads to the matrix equation
written compactly as
\begin{eqnarray}
\nonumber
\les\begin{array}{c}
A_{smn}^{(1)}\\[5pt]
-  -  -  -  \\[5pt]
B_{smn}^{(1)}
\end{array}\ris
&=&
\les\begin{array}{c|c}
I_{smn,{s^\prime}{m^\prime}{n^\prime}}^{(1)} &
J_{smn,{s^\prime}{m^\prime}{n^\prime}}^{(1)}\\[5pt]
-  -  -  -  -  -  & -  -  -  -  -  -  \\[5pt]
K_{smn,{s^\prime}{m^\prime}{n^\prime}}^{(1)} &
L_{smn,{s^\prime}{m^\prime}{n^\prime}}^{(1)}
\end{array}\ris
\les\begin{array}{c}
b_{{s^\prime}{m^\prime}{n^\prime}}^{(1)}\\[5pt]
-  -  -  -  \\[5pt]
c_{{s^\prime}{m^\prime}{n^\prime}}^{(1)}
 \end{array}\ris
 \\[8pt]
 \label{Y1}
 &\equiv&
 \les {Y^{(1)}}\ris
 \les\begin{array}{c}
b_{{s^\prime}{m^\prime}{n^\prime}}^{(1)}\\[5pt]
-  -  -  -  \\[5pt]
c_{{s^\prime}{m^\prime}{n^\prime}}^{(1)}
 \end{array}\ris\,;
\end{eqnarray}
likewise, Eqs.~\r{Asmn3}--\r{Hint} yield
written compactly as
\begin{eqnarray}
\nonumber
\les\begin{array}{c}
A_{smn}^{(3)}\\[5pt]
-  -  -  -  \\[5pt]
B_{smn}^{(3)}
\end{array}\ris
&=&-\,
\les\begin{array}{c|c}
I_{smn,{s^\prime}{m^\prime}{n^\prime}}^{(3)} &
J_{smn,{s^\prime}{m^\prime}{n^\prime}}^{(3)}\\[5pt]
-  -  -  -  -  -  & -  -  -  -  -  -  \\[5pt]
K_{smn,{s^\prime}{m^\prime}{n^\prime}}^{(3)} &
L_{smn,{s^\prime}{m^\prime}{n^\prime}}^{(3)}
\end{array}\ris
\les\begin{array}{c}
b_{{s^\prime}{m^\prime}{n^\prime}}^{(1)}\\[5pt]
-  -  -  -  \\[5pt]
c_{{s^\prime}{m^\prime}{n^\prime}}^{(1)}
 \end{array}\ris
 \\[8pt]
 \label{Y3}
 &\equiv&-\,
 \les {Y^{(3)}}\ris
 \les\begin{array}{c}
b_{{s^\prime}{m^\prime}{n^\prime}}^{(1)}\\[5pt]
-  -  -  -  \\[5pt]
c_{{s^\prime}{m^\prime}{n^\prime}}^{(1)}
 \end{array}\ris\,.
\end{eqnarray}
These relations require the computation of the integrals
\begin{eqnarray}
&&\nonumber
I_{smn,{s^\prime}{m^\prime}{n^\prime}}^{(j)}= -\frac{i\ko^2}{\pi}
\int_\sfS d^2{\#r_p}\left( \#N\smn^{(\ell)}(\ko\#r_p)\.\lec
 \#n(\#r_p)\times\les  \=B\.\newM_{{s^\prime}{m^\prime}{n^\prime}}^{(1)}(\#r_p)\ris\ric
\right.
\\
&&\qquad\left.
+ \frac{1}{\eta_r}\#M\smn^{(\ell)}(\ko\#r_p)\.\lec  \#n(\#r_p)\times \les
\=B\. \newN_{{s^\prime}{m^\prime}{n^\prime}}^{(1)}(\#r_p)\ris
\ric\right)\,,
\\
&&\nonumber
J_{smn,{s^\prime}{m^\prime}{n^\prime}}^{(j)}= -\frac{i\ko^2}{\pi}
\int_\sfS d^2{\#r_p}\left( \#N\smn^{(\ell)}(\ko\#r_p)\.\lec
 \#n(\#r_p)\times\les  \=B\.\newN_{{s^\prime}{m^\prime}{n^\prime}}^{(1)}(\#r_p)\ris\ric
\right.
\\
&&\qquad\left.
+ \frac{1}{\eta_r}\#M\smn^{(\ell)}(\ko\#r_p)\.\lec  \#n(\#r_p)\times \les
\=B\. \newM_{{s^\prime}{m^\prime}{n^\prime}}^{(1)}(\#r_p)\ris
\ric\right)\,,
\\
&&\nonumber
K_{smn,{s^\prime}{m^\prime}{n^\prime}}^{(j)}= -\frac{i\ko^2}{\pi}
\int_\sfS d^2{\#r_p}\left( \#M\smn^{(\ell)}(\ko\#r_p)\.\lec
 \#n(\#r_p)\times\les  \=B\.\newM_{{s^\prime}{m^\prime}{n^\prime}}^{(1)}(\#r_p)\ris\ric
\right.
\\
&&\qquad\left.
+ \frac{1}{\eta_r}\#N\smn^{(\ell)}(\ko\#r_p)\.\lec  \#n(\#r_p)\times \les
\=B\. \newN_{{s^\prime}{m^\prime}{n^\prime}}^{(1)}(\#r_p)\ris
\ric\right)\,,
\\
&&\nonumber
L_{smn,{s^\prime}{m^\prime}{n^\prime}}^{(j)}= -\frac{i\ko^2}{\pi}
\int_\sfS d^2{\#r_p}\left( \#M\smn^{(\ell)}(\ko\#r_p)\.\lec
 \#n(\#r_p)\times\les  \=B\.\newN_{{s^\prime}{m^\prime}{n^\prime}}^{(1)}(\#r_p)\ris\ric
\right.
\\
&&\qquad\left.
+ \frac{1}{\eta_r}\#N\smn^{(\ell)}(\ko\#r_p)\.\lec  \#n(\#r_p)\times \les
\=B\. \newM_{{s^\prime}{m^\prime}{n^\prime}}^{(1)}(\#r_p)\ris
\ric\right)\,,
\end{eqnarray}
where $j\in\lec1,3\ric$ and $\ell = j+2\, (\mbox{mod} \,4)\in\lec3,1\ric$.

Equations \r{Y1} and \r{Y3} yield a matrix relationship between the incident-field coefficients
and the scattered-field coefficients that can be expressed as
\begin{equation}
\les\begin{array}{c}
A_{smn}^{(3)}\\[5pt]
-  -  -  -  \\[5pt]
B_{smn}^{(3)}
\end{array}\ris
= \les {T}\ris
\les\begin{array}{c}
A_{smn}^{(1)}\\[5pt]
-  -  -  -  \\[5pt]
B_{smn}^{(1)}
\end{array}\ris\,,
\end{equation}
where the T matrix
\begin{equation}
\les {T}\ris= -\,
 \les {Y^{(3)}}\ris
 \les {Y^{(1)}}\ris^{-1}
 \end{equation}
 completely characterizes the scattering response of the 3D anisotropic object.

 The calculation of the T matrix of a 3D object made of a material obeying the constitutive
 relations \r{con-D} and \r{con-B} along with \r{stillgen} is a major exercise that we intend to take up shortly. Let us, however, mention
 that our derivation procedure
 can be extended to cover the possibility that the external region
 is occupied by a homogeneous isotropic material which can be dielectric-magnetic,
 chiral, or even biisotropic \cite{Lakhtakia1992}.

 \section{Radiation fields}\label{radfields}
 Returning to Eqs.~\r{Faraday1} and \r{Ampere1}, we are now interested in finding the electric field radiated by a source current density
 distribution. Accordingly,
 \begin{equation}
 \label{EroG}
 \#E\ro=i\omega\muo\mur\int\int\int d^3\#r_p \, \=G(\#r,\#r_p)\.\#J(\#r_p)\,,
 \end{equation}
 where $\=G(\#r,\#r_p)$ is the dyadic Green function with $\#r$ as the field point and $\#r_p$ as the source point.

 \subsection{Dyadic Green function: Coordinate-free form}\label{sec:dgf}
Our starting point is the particular solution of Eq.~\r{Helmholtz4e}, which is known to be \cite{Chen}
 \begin{equation}
 \label{etog}
 \~e(\#r^\prime)=i\omega\muo\mur\,A \int\int\int d^3\#r^\prime_p \, \tilde{\=g}(\#r^\prime,\#r^\prime_p)\.\~j(\#r^\prime_p)\,,
 \end{equation}
where
\begin{equation}
\label{gto}
\tilde{\=g}(\#r^\prime,\#r^\prime_p) =  \left(\=I +k^{-2} {\nabla^\prime\nabla^\prime}\right)
\frac{\exp\le{ik\vert\#r^\prime-\#r^\prime_p\vert}\ri}{4\pi\vert\#r^\prime-\#r^\prime_p\vert}
\end{equation}

Inverting the relations \r{erto} and \r{jrto}, we get
 \begin{equation}
 \label{epog}
 \#e^\prime(\#r^\prime)=i\omega\muo\mur  \int\int\int d^3\#r^\prime_p \,  {\=g}^\prime(\#r^\prime,\#r^\prime_p)\.\#j^\prime(\#r^\prime_p)\,
 \end{equation}
 from Eq.~\r{etog}, where
\begin{eqnarray}
\label{gpo}
{\=g}^\prime(\#r^\prime,\#r^\prime_p) &=& A^3\=A^{-1}\.\tilde{\=g}(\=A^{-1}\.\#r^\prime,\=A^{-1}\.\#r^\prime_p)\.\=A^{-1}
\\
&=& A^3\=A^{-1}\.\left(\=I +k^{-2} \=A\.{\nabla^\prime\nabla^\prime}\.\=A\right)\.\=A^{-1}\,
\frac{\exp\les{ik\vert\=A^{-1}\.\le\#r^\prime-\#r^\prime_p\ri\vert}\ris}{4\pi\vert\=A^{-1}\.\le\#r^\prime-\#r^\prime_p\ri\vert}\\
\label{gpo-final}
&=& A^3 \left(\=A^{-2} +k^{-2}  {\nabla^\prime\nabla^\prime} \right) \,
\frac{\exp\les{ik\vert\=A^{-1}\.\le\#r^\prime-\#r^\prime_p\ri\vert}\ris}{4\pi\vert\=A^{-1}\.\le\#r^\prime-\#r^\prime_p\ri\vert}\,,
\end{eqnarray}
and the identity $\=A^T = \=A$ has been used.

Next, an inversion involving Eqs.~\r{erpo} and \r{jrpo}  transforms Eq.~\r{epog} to
 \begin{equation}
 \label{erog}
 \#e(\#r)=i\omega\muo\mur  \int\int\int d^3\#r_p \,  {\=g}(\#r,\#r_p)\.\#j(\#r_p)\,,
 \end{equation}
 where
 \begin{eqnarray}
 \label{gro}
{\=g} (\#r,\#r_p) &=& \=S\.{\=g}^\prime(\=S\.\#r,\=S\.\#r_p)\.\=S^{-1}
\\
&=&
A^3\=S\. \left(\=A^{-2} +k^{-2}  \=S\.{\nabla\nabla}\.\=S^{-1} \right)\. \=S^{-1}
\frac{\exp\les{ik\vert\=A^{-1}\.\=S\.\le\#r-\#r_p\ri\vert}\ris}{4\pi\vert\=A^{-1}\.\=S\.\le\#r-\#r_p\ri\vert}
\\
\label{gro-final} &=& A^3  \left(\=S\.\=A^{-2}\. \=S^{-1} +k^{-2}
\=S^2\.{\nabla\nabla}\.\=S^{-2} \right)
\frac{\exp\les{ik\vert\=A^{-1}\.\=S\.\le\#r-\#r_p\ri\vert}\ris}{4\pi\vert\=A^{-1}\.\=S\.\le\#r-\#r_p\ri\vert}.
\end{eqnarray}

Finally, the use of Eqs.~\r{ero} and \r{jro} in Eq.~\r{erog} yields
Eq.~\r{EroG} with
\begin{eqnarray}
 \label{Gro}
{\=G} (\#r,\#r_p)&=& \exp\les{i\ko\#\Gamma}\.\le\#r-\#r_p\ri\ris{\=g} (\#r,\#r_p)\,
\\
\nonumber
&=&
A^3 \exp\les{i\ko\#\Gamma}\.\le\#r-\#r_p\ri\ris \left(\=S\.\=A^{-2}\. \=S^{-1} \right.
 \\
 \label{Gro-final}
 &&\quad\left.
 +k^{-2}  \=S^2\.{\nabla\nabla}\.\=S^{-2} \right)
 \frac
 {\exp\les{ik\vert\=A^{-1}\.\=S\.\le\#r-\#r_p\ri\vert}\ris}
 {4\pi{\vert\=A^{-1}\.\=S\.\le\#r-\#r_p\ri\vert}}\,
 \end{eqnarray}
 in a coordinate-free form.

\subsection{Dyadic Green function: Singularity}
When the
source point $\#r_p$ and the field point $\#r$ coincide, the dyadic Green function
exhibits a singularity.
In order to estimate this singularity,
the source point is made the centroid of an
electrically small convex region $\sfVl$ bounded by the surface
$\sfSl$, which is then shrunk in volume isotropically
\cite{Yaghjian1980}. This process is used to determine the
depolarization dyadic $\=D$ defined from  $\=G(\#r,\#r_p)$  via
\begin{equation}
\label{depol-def}
\lim_{\ell \to 0}
\int\int\int_{\sfVl} d^3\#r_q\, \=G(\#r_p,\#r_q)\.\#b(\#r_q)\simeq \=D\.\#b(\#r_p)\,,
\end{equation}
where the surface
\begin{equation}
\sfSl = \lec \#r_q \Big\vert \#r_q(\Omega_\ell) = \#r_p + \ell \Ul\.{\#u}_\ell(\Omega_\ell),\, \Omega_\ell\in\les0,4\pi\ris\ric\,.
\end{equation}
Here, the positive scalar $\ell$ is a linear measure of the volume of $\sfVl$; the dyadic $\Ul$ delineating the shape
of $\sfVl$ has a determinant equal to unity, while ${\#u}_\ell$ and $\Omega_\ell$ denote, respectively, the radial unit vector and the solid angle in a spherical coordinate system with its origin at $\#r_p\in\sfVl$. The   depolarization dyadic $\=D$  depends on the choice of $\Ul$ \cite{Yaghjian1980}.

 In the present instance, regardless of the choice of $\Ul$, two
different approaches \cite{LMWaeu} to determining $\=D$ are possible
as follows:
\begin{itemize}
\item[(a)] Determine the analogous depolarization dyadic $\tilde{\=d}$ arising from $\tilde{\=g}(\#r^\prime,\#r^\prime_p)$, and then reverse the three
transformations described in
Secs.~\ref{1st_transformation}--\ref{3rd_transformation} to obtain $\=D$.
Implementing this indirect approach, we find that Eq.~\r{gto}
leads to
\begin{equation}
\label{gto-sing}
\tilde{\=d}=- k^{-2} \=L\,,
\end{equation}
where
\begin{equation}
\=L=\int\int_{\sfSl}\,d^2\#r_q \frac{\#u_q\#r_q-\#u_q\#r_p}{4\pi\vert{\#r_q-\#r_p}\vert^3}
\end{equation}
and $\#u_q$ is the unit outward normal to $\sfSl$ at
$\#r_q\in\sfSl$. The dyadic $\=L$ is symmetric \cite{Yaghjian1980}. If $\sfVl$ is either spherical or cubical, $\=L=(1/3)\=I$. For ellipsoidal $\sfVl$, the dyadic
$\=L$ can be cast in terms of elliptic integrals that have to be evaluated numerically \cite{Osborn1945,Stoner1945}. Using Eqs.~\r{gpo} and \r{gto-sing}, we get
the
\begin{equation}
\label{gpo-sing}
{\=d}^\prime=- k^{-2}A^3\=A^{-1}\.\=L\.\=A^{-1}\,
\end{equation}
arising from ${\=g}^\prime(\#r^\prime,\#r^\prime_p)$, and finally
\begin{equation}
\label{Gro-sing}
{\=D}=- k^{-2}A^3\=S\.\=A^{-1}\.\=L\.\=A^{-1}\.\=S^{-1}\,
\end{equation}
emerges from the use of Eqs.~\r{gro} and \r{Gro}.

\item[(b)]  Directly obtain $\=D$ for the material described by the
constitutive relations \r{con-D} and \r{con-B}, using the method of
Michel and Weiglhofer \c{MW}.
This direct approach was formulated for ellipsoidal $\sfVl$ and
yields \c{MW} 
\begin{eqnarray}
\=D &=&  - \frac{A^2}{4 \pi  k^2 }
\int_{\varphi_\rho=0}^{2\pi}\,d\varphi_\rho
\int_{\vartheta_\rho=0}^{\pi} \,d\vartheta_\rho\,\sin\vartheta_\rho
\frac{\Ul^{-1}\.\urho\urho\.\Ul^{-1}}
{\urho\.\Ul^{-1}\.\=S\.\=A\.\=A\.\=S^{-1}\.\Ul^{-1}\.\urho}
\,,  \l{depol}
\end{eqnarray}
where $\urho=\le\ux\cos\varphi_\rho+\uy\sin\varphi_\rho\ri\sin\vartheta_\rho+\uz\sin\vartheta_\rho$.

\end{itemize}

For spherical $\sfVl$, the right side of Eq.~\r{Gro-sing} simplifies to $-(A^3/3k^2)\=S\.\=A^2\.\=S$,
but the integral on the right side Eq.~\r{depol}  has still to be expressed in terms of
elliptic integrals  \c{W}.
Therefore,  the depolarization dyadics  \r{Gro-sing}
and \r{depol}---obtained by two different approaches---are clearly distinct in general. The difference arises
because, for the two approaches, 
\begin{itemize}
\item[(i)] $\sfVl$ was chosen in different spaces: $\#r^\prime$ in the indirect approach and $\#r$ in the
direct approach;
 and 
 \item[(ii)] the singularity was
considered for different fields $\~e$ in the indirect approach and $\#e$ (or $\#E$) in the direct approach.
\end{itemize}
Knowledge of the depolarization dyadic will be useful in setting up the Maxwell Garnett and the Bruggeman
formalisms to predict the effective constitutive parameters of a particulate composite material containing
electrically small particles made of the chosen material \cite{WLMmotl}.

\subsection{Dyadic Green function: Bilinear expansion}\label{bilinear}
Computation of the field radiated by a source can be greatly convenienced by the formulation of a blinear expression of 
 the dyadic Green function $ \=G(\#r,\#r_p)$, because
the bilinear expression  is a sum over the products of the derived vector spherical wavefunctions
of the source point and the field point \cite{Wood,LIap}.
It is applicable only if the source point and the field point are situated at different distances from the origin.

We begin with the known bilinear expansion \cite{BY1975,LVV1985}
 \begin{equation}
 \label{gto1}
\tilde{\=g}(\#r^\prime,\#r^\prime_p)=\frac{ik}{\pi}\sum_{s=e,o}\sum_{n=1}^{\infty}\sum_{m=0}^{n}\,
D_{mn} \les \#M\smn^{(3)}(k\#r^\prime_>)\#M\smn^{(1)}(k\#r^\prime_<) + \#N\smn^{(3)}(k\#r^\prime_>)\#N\smn^{(1)}(k\#r^\prime_<)\ris\,,
\end{equation}
where
\begin{equation}
\#r^\prime_> =\lec
\begin{array}{ll}
\#r^\prime\,, &r^\prime>r^\prime_p\\
\#r^\prime_p\,, &r^\prime<r^\prime_p
\end{array}\right.\,,\qquad\quad
\#r^\prime_< =\lec
\begin{array}{ll}
\#r^\prime_p\,, &r^\prime>r^\prime_p\\
\#r^\prime\,, &r^\prime<r^\prime_p
\end{array}\right.\,.
\end{equation}
By following the process described in Sec.~\ref{sec:dgf}, we get
\begin{eqnarray}
\nonumber
&&{\=g}^\prime(\#r^\prime,\#r^\prime_p)=\frac{ikA^3}{\pi}
\=A^{-1}\.
\sum_{s=e,o}\sum_{n=1}^{\infty}\sum_{m=0}^{n}\,
D_{mn} \left[ \#M\smn^{(3)}(k\=A^{-1}\.\#r^\prime_>)\#M\smn^{(1)}(k\=A^{-1}\.\#r^\prime_<) \right.
\\
 \label{gpo1}
 &&\qquad\qquad\left.
 + \#N\smn^{(3)}(k\=A^{-1}\.\#r^\prime_>)\#N\smn^{(1)}(k\=A^{-1}\.\#r^\prime_<)\right]
\.\=A^{-1}
\,,
\end{eqnarray}
and
\begin{eqnarray}
\nonumber
&&{\=g}(\#r,\#r_p)=\frac{ikA^3}{\pi}\exp\les{-i\ko\#\Gamma\.\le\#r_>+\#r_<\ri}\ris
\\
 \label{gro1}
 &&\qquad\qquad\times
 \sum_{s=e,o}\sum_{n=1}^{\infty}\sum_{m=0}^{n}\,
D_{mn} \left[ \#m\smn^{(3)}(\#r_>)\#m\smn^{(1)}(\#r_<)
 + \#n\smn^{(3)}(\#r_>)\#n\smn^{(1)}(\#r_<)\right]
\,,
\end{eqnarray}
where
\begin{equation}
\#r_> =\lec
\begin{array}{ll}
\#r\,, &r>r_p\\
\#r_p\,, &r<r_p
\end{array}\right.\,,\qquad\quad
\#r_< =\lec
\begin{array}{ll}
\#r_p\,, &r>r_p\\
\#r\,, &r<r_p
\end{array}\right.\,.
\end{equation}

Finally, from Eqs.~\r{Gro} and \r{gro1}, we get the desired bilinear expansion
\begin{eqnarray}
\nonumber
&&{\=G}(\#r,\#r_p)=\frac{ikA^3}{\pi}\exp\les{-i\ko\#\Gamma\.\le\#r_>+\#r_<-\#r+\#r_p\ri}\ris
\\
 \label{Gro1}
 &&\qquad \times
 \sum_{s=e,o}\sum_{n=1}^{\infty}\sum_{m=0}^{n}\,
D_{mn} \left[ \#m\smn^{(3)}(\#r_>)\#m\smn^{(1)}(\#r_<)
 + \#n\smn^{(3)}(\#r_>)\#n\smn^{(1)}(\#r_<)\right]
\,.
\end{eqnarray}
Application of the conditions \r{stillgen} simplifies the foregoing
expression as follows:
\begin{eqnarray}
\nonumber
&&{\=G}(\#r,\#r_p)=\frac{ikA^3}{\pi}\exp\les{-i\ko\#\Gamma\.\le\#r_>+\#r_<-\#r+\#r_p\ri}\ris
\\
 \label{Gro2}
 &&\qquad \times\=B\.
 \sum_{s=e,o}\sum_{n=1}^{\infty}\sum_{m=0}^{n}\,
D_{mn} \left[ \newM\smn^{(3)}(\#r_>)\newM\smn^{(1)}(\#r_<)
 + \newN\smn^{(3)}(\#r_>)\newN\smn^{(1)}(\#r_<)\right]\.\=B
\,.
\end{eqnarray}
Equations \r{Gro1} and \r{Gro2} will be helpful for analyzing  aperture antennas \cite{Wood,LIap}
and radomes made of the chosen material \cite{EK,LVVjpd}.

\section{Concluding remarks}

The derivation, in closed-form, of vector spherical wavefunctions
for an orthorhombic dielectric-magnetic material with
gyrotropic-like magnetoelectric properties, enables the
electromagnetic scattering response of an object made of such a material to
be comprehensively and  conveniently characterized. In particular,
these wavefunctions provide the building blocks for (i) the T-matrix
formulation which yields a description of the field scattered by a
3D object composed of the chosen material, and (ii) the dyadic Green
function which yields a description of the radiation field of a
source current density distribution embedded in the chosen material.\\

\noindent\red{ {\bf Acknowledgments.}
AL thanks Craig F. Bohren for a discussion regarding the antecedents of the 1908 paper
of Mie \cite{Mie1908}, and the Charles Godfrey Binder Endowment at Penn State for ongoing
support of his research activities.}

\end{document}